\documentclass[aps,prl,reprint,superscriptaddress]{revtex4-1}
\pdfoutput=1
\usepackage{amsmath}
\usepackage{amssymb}
\usepackage{graphicx}
\usepackage{dcolumn}
\usepackage{bm}
\usepackage{color}
\usepackage{natbib}
\begin{document}

\title{ Quantum-dot states and optical excitations in edge-modulated 
          graphene nanoribbons }
\author{Deborah \surname{Prezzi}}
\email[Corresponding author: ]{deborah.prezzi@unimore.it}
\affiliation{Centro S3, CNR-Istituto Nanoscienze, Via G. Campi 213/A, 
             I-41125 Modena, Italy}
\author{Daniele \surname{Varsano}}
\altaffiliation[Present address: ]{Dipartimento di Fisica, Universit\`a 
                di Roma ``La Sapienza'', P.le Aldo Moro 2, I-00185 Roma, 
                Italy}
\affiliation{Centro S3, CNR-Istituto Nanoscienze, Via G. Campi 213/A, 
             I-41125 Modena, Italy}
\author{Alice \surname{Ruini}}
\affiliation{Centro S3, CNR-Istituto Nanoscienze, Via G. Campi 213/A, 
             I-41125 Modena, Italy}
\affiliation{Dipartimento di Fisica, Universit\`a di Modena e Reggio 
             Emilia, via G. Campi 213/A, I-41125 Modena, Italy}
\author{Elisa \surname{Molinari}}
\affiliation{Centro S3, CNR-Istituto Nanoscienze, Via G. Campi 213/A, 
             I-41125 Modena, Italy}
\affiliation{Dipartimento di Fisica, Universit\`a di Modena e Reggio 
             Emilia, via G. Campi 213/A, I-41125 Modena, Italy}
\date{\today}

\begin{abstract}
\noindent
We investigate from first principles the electronic and optical properties of 
edge-modulated armchair graphene nanoribbons, including both quasi-particle 
corrections and excitonic effects.
Exploiting the oscillating behavior of the ribbon energy gap, we show that 
minimal width modulations are sufficient to obtain confinement of both 
electrons and holes, thus forming optically active quantum dots 
with unique properties, such as cohexistence of dot-like and extended 
excitations and fine tunability of optical spectra, with great potential for
optoelectronic applications.
\end{abstract}

\maketitle


Graphene nanostructures have recently triggered a wealth of studies for their 
remarkable properties, which combine the unique electronic and mechanical 
features of graphene~\cite{cast+09rmp,*geim09sci} with the semiconducting 
behavior induced by quantum confinement~\cite{han+07prl,*chen+07pe}. 
Moreover, depending on the details of the atomic structure, a variety of 
novel width and edge-related phenomena can arise~\cite{naka+96prb,*son+06prl}. 
To fully exploit this richness, great efforts have been devoted to achieve 
precise control of the structure through a number of different 
nanofabrication techniques \cite{[{For review, see e.g. }][{}]wei-liu10am,*roch11natn}.
Extreme control of the width --down to three benzene rings-- was recently 
demonstrated by chemical routes for armchair-edge graphene nanoribbons (A-GNRs) 
\cite{cai+10nat}, whose optical gaps \cite{prez+07pssb,prez+08prb,%
yang+07nl,cocc+11tobe} finally reach the energy window attractive for 
optoelectronic applications.
The perspective of engineering 1D and 0D quantum confinement of charge 
carriers, so far dominated by unintentional disorder effects 
\cite{stam+09prl,*gall+10prb,*han+10prl}, is thus becoming realistic 
\cite{shim+11natn}.

Among the different routes proposed to achieve quantum confinement in all 
directions \cite{[{See e.g. }]silv-efet07prl,*wang+07apl,*sevi+08prb,
*borc+10prl,*sing+10nano}, i.e. quantum dots (QDs), the most widely used 
takes advantage of high-resolution lithography to carve the full device from 
2D graphene \cite{pono+08sci}.
This allows to combine atomic-like properties --extensively explored in 
conventional semiconductor QDs \cite{ross05book}-- with the advantages born 
by graphene, such as efficient coupling to a graphene-based interconnecting 
wire or contact and planar geometry compatible with available technologies. In 
such a configuration, both dot and barriers are made of graphene by modulating 
the wire width, and the carriers result to be localized in the wider region by 
quantum size effects, as in conventional semiconductor nanostructures.

\begin{figure}[t!]
\includegraphics[width=.40\textwidth,clip]{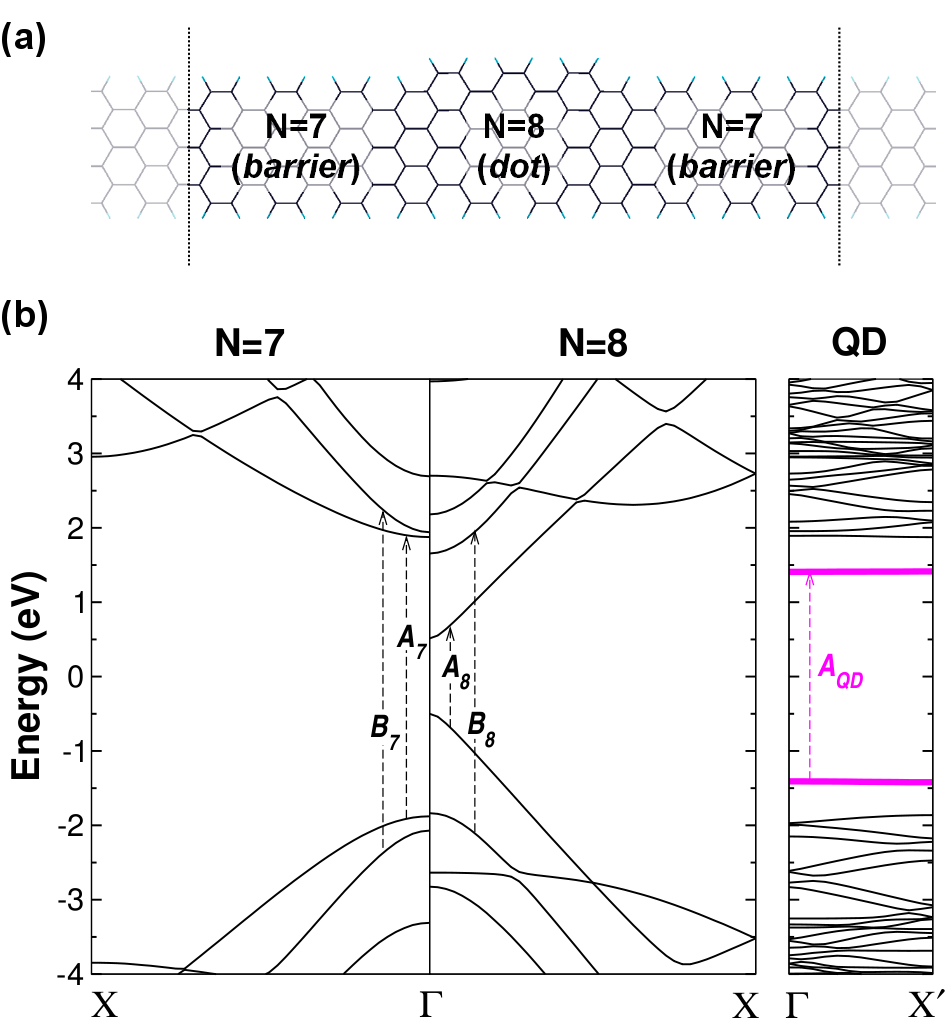}
\caption{
(Color online) {\bf (a)} Atomic structure of the modulated graphene nanoribbon. 
The dot and the barrier regions ($N=8$ and $N=7$, respectively) differ in width
by one atomic row. The armchair edges are H-terminated. 
{\bf (b)} Quasi particle band dispersions for the infinite GNRs $N=7$ and 
$N=8$ (left panels), and for the periodically repeated QD (right panel), whose
unit cell is  highlighted in (a) by dotted lines. The two non dispersive 
states (thick magenta lines) are localized in the dot ($N=8$) region of the 
structure. Arrows indicate the lowest optically allowed transitions (see 
Fig.~\ref{EXCeps}).
}\label{QDstruct-band}
\end{figure}

Here we show that not only quantum size effects, but also a novel 
mechanism --similar to that occurring in conventional semiconductor 
{\it hetero}junctions-- can give rise to confinement of both electrons and 
holes, where the the confining potential landscape is obtained by exploiting 
the peculiar electronic properties of AGNRs. 
In fact, the energy band gap of AGNRs shows three distinct families depending
on the ribbon width, namely $N=3p$, $N=3p+1$, and $N=3p+2$ ($N$ being the 
number of dimeric lines along the width, and $p$ a positive integer)
\cite{son+06prl}: Within each family, the energy gap decreases with increasing 
ribbon width, as expected, but minimal width modulations, down to one atomic 
row, are sufficient to induce a large variation in the band gap.
For example, the conduction and the valence band offsets between the $N=7$ 
and the $N=8$ ribbon are both as large as 1.4 eV, as will be discussed 
below (see Fig.~\ref{QDstruct-band}).

Joining such ribbons is therefore expected to produce an all-graphene
system with type-I band modulation similar to a semiconductor interface
between different materials, while at the same time retaining graphene
$\pi$-conjugation throughout. In this paper, we provide a realistic description 
of QD states and optical excitations for such edge-modulated AGNRs by means 
of state-of-the-art ab-initio approaches. 
Many-body effects, which are known to dominate electronic and optical 
properties in low-dimensional systems~\cite{prez-moli06pssa,prez+08prb,
ruin+02prl,chan+04prl,*maul+05prb}, are also included via the $GW$ and 
Bethe-Salpeter (BS) schemes \cite{mbpt-review}. These solid-state techniques, 
though considerably heavy in our case, are crucial to 
study both dot-like and extended features on equal footing, thus allowing us to
obtain quantitative predictions of the optical spectra. 
Our whole results demonstrate that edge-modulated AGNRs offer a novel mechanism 
for the creation of optically active carbon-based QDs with prominent and 
tunable exciton localization features, which make them suitable for a variety
of applications, ranging from single-photon emission to optically-driven 
quantum information. 


As displayed in Fig.~\ref{QDstruct-band}, we build our prototype graphene-based 
QD by considering a superlattice obtained from the periodically repeated 
junction of two AGNRs belonging to two different families, $N=7$ 
({\it barrier}) and $N=8$ ({\it dot}), so as to {\it maximize} the energy gap 
difference and {\it minimize} the width variation. 
Once the two constituents of the superlattice are defined, the {\it length } 
of both barrier and dot region will then determine the depth and the number 
of quantum-dot states according to the confinement mechanism described above. 
Here we chose these parameters in a way to guarantee the presence of a couple 
of localized states maintaining the feasibility of calculations
\footnote{The length of the $N=7$ ($N=8$) segment is about 3 nm (1.3 nm). 
For shorter barrier length (e.g. $\sim 1.7$ nm) the first states in valence 
and conduction band are no more localized; enlarging the dot region (e.g. 
from 1.3 to 2.1 nm) makes instead the localized states deeper in energy.}.  
For comparison, $N=7$ and $N=8$ ideal GNRs are also studied. In all systems, 
dangling bonds at the edges are saturated with monoatomic hydrogen 
\cite{wass+08prl}.

The systems described above were fully relaxed by performing density-functional 
theory (DFT) supercell calculations within the local density approximation 
(LDA), as implemented in the {\sc Quantum ESPRESSO} package \cite{QEspresso,epaps}. 
In order to improve the band structure description obtained at the DFT-LDA 
level, we then computed the quasi-particle corrections to the Kohn-Sham 
eigenvalues within the $G_0W_0$ approximation for the self-energy operator. 
In addition, excitonic effects were taken into account by solving the Bethe-Salpeter 
(BS) equation, which describes the exciton dynamics in terms of the screened 
quasielectron-quasihole interaction \cite{mbpt-review}.
From the solution of the BS equation, the absorption spectra were then computed
as the imaginary part of the macroscopic dielectric function.
The inclusion of the aforementioned many-body effects was done using the 
{\sc Yambo} code~\cite{mari+09cpc,epaps}.

%
%
%

Figure~\ref{QDstruct-band}(b) depicts the quasi-particle band structures of 
the studied systems, that is $N=7$, $N=8$, and the modulated GNR superlattice.
In the superlattice (right panel), the band gap difference between $N=7$ and 
$N=8$ acts as a confining potential for the $N=8$ region, giving rise to an 
effective one-dimensional potential well (periodically repeated). 
This confining potential is sufficient to localize two states in the well 
region (see thick non-dispersive bands in magenta), which thus behaves as a 
quantum dot for both electrons and holes. 
As found for ideal GNRs~\cite{prez+08prb}, the quasi-particle corrections to 
the DFT-LDA energy gap are particularly large if compared with standard 
semiconductors, due to a much weaker screening and the quasi one-dimensional 
nature of the systems, which both concur to enhance the electron-electron 
interaction. This effect is evident also for the superlattice structure, where 
the LDA energy gap between the localized states is increased from 1.0 to 2.8 
eV, that is, the $G_0W_0$ energy gap is almost three times the LDA one.
In addition, both ideal and modulated GNRs show an overall stretching of the 
band structure of about 20\%. 

%
%
\begin{figure}
\includegraphics[width=.33\textwidth,clip]{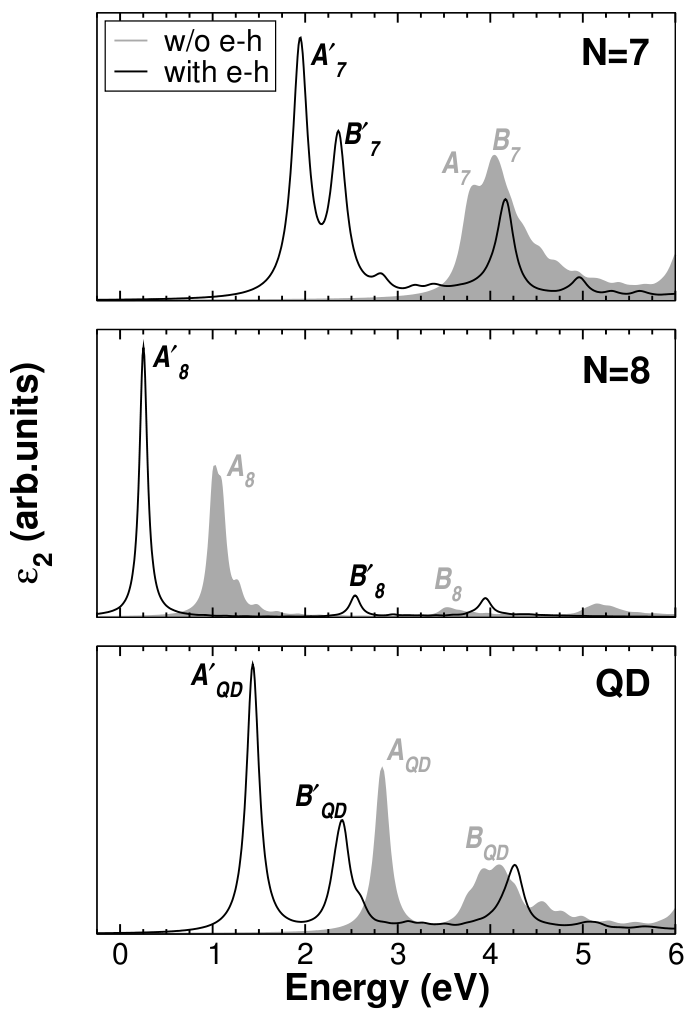}
\caption{Optical absorption spectra of $N=7$, $N=8$, and the modulated GNR (QD).
The solid black line represents the spectrum with the inclusion of the 
electron-hole interaction, while the $GW$-RPA one (i.e. without e-h interaction)
is in grey. 
All the spectra are computed for light polarization parallel to the ribbon axis,
introducing a Lorentzian broadening of 0.08 eV.
}\label{EXCeps}
\end{figure}

The optical absorption spectra for the superlattice and the two subsystems 
are reported in Fig.~\ref{EXCeps}, as obtained both within the
random-phase approximation (i.e. $GW$-RPA, grey shaded area) and including 
electron-hole interaction (black line). The main transitions giving rise to 
the low-energy peaks are indicated with vertical arrows in Fig.~\ref{QDstruct-band}.
As can be seen, the inclusion of excitonic effects, which dramatically modifies 
both peak position and absorption lineshape, is crucial to give both 
qualitative and quantitative predictions of all the optical spectra. The 
prominent 1D van Hove singularities characterizing the RPA spectra disappear 
giving rise to individual excitonic states below the onset of the continuum, 
with exceedingly large binding energy. This produces an overall red-shift 
of the spectrum, as opposite to the $GW$ gap opening, together with a change 
in the relative position of the peaks and possibly in their relative intensity.   

In ideal ribbons, the position of the first peak exhibits an oscillating 
behaviour, according to the family classification described above 
\cite{prez+08prb}. This allows to span an energy window of more than 1.5 eV 
just by changing the width of one atomic row. In the case of the superlattice, 
a peak arises between $A_7$ and $A_8$, which corresponds to the interband 
transition between the states localized in the dot region within the RPA 
picture [$A_{QD}$, see Fig.~\ref{QDstruct-band}(b), right panel]. 
As previously mentioned, the structural parameters of the superlattice 
crucially determine the number and energy location of the confined states, 
that is, the relative peak position wrt $A_7$ and $A_8$. This would further 
improves the flexibility of this class of systems, where a fine tuning of 
the QD spectrum can be enabled both by interfacing different GNRs and by 
changing the dot/barrier length within a given superstructure. 

\begin{figure}
\includegraphics[width=.34\textwidth,clip]{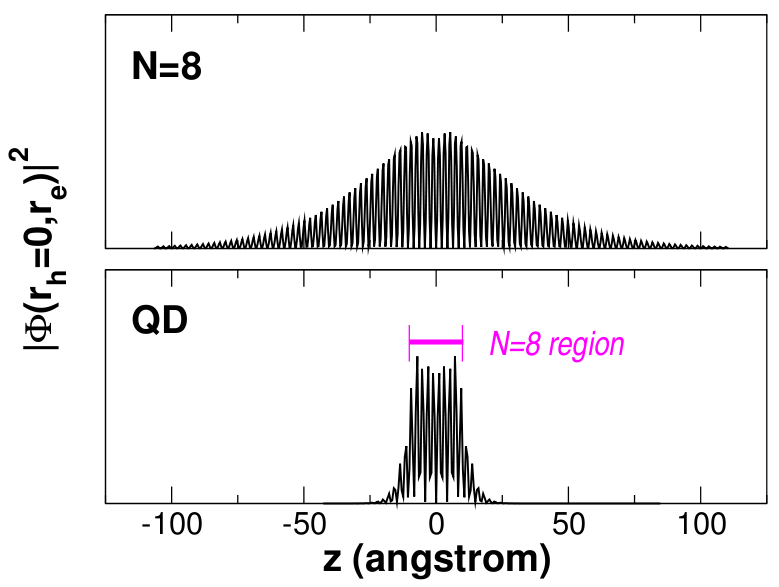}
\caption{Electron distribution of the lowest excitons for $N=8$ and the 
edge-modulated GNR (QD), for a fixed hole position (here set to $r_h=0$). 
The distribution is plotted along the ribbon axis ($z$), after integrating out 
the other coordinates. For the superlattice $r_h$ is chosen to be at the 
center of the dot ($N=8$) region.
}\label{EXCwfc1d}
\end{figure}

\begin{figure}[b!]
\includegraphics[width=.47\textwidth,clip]{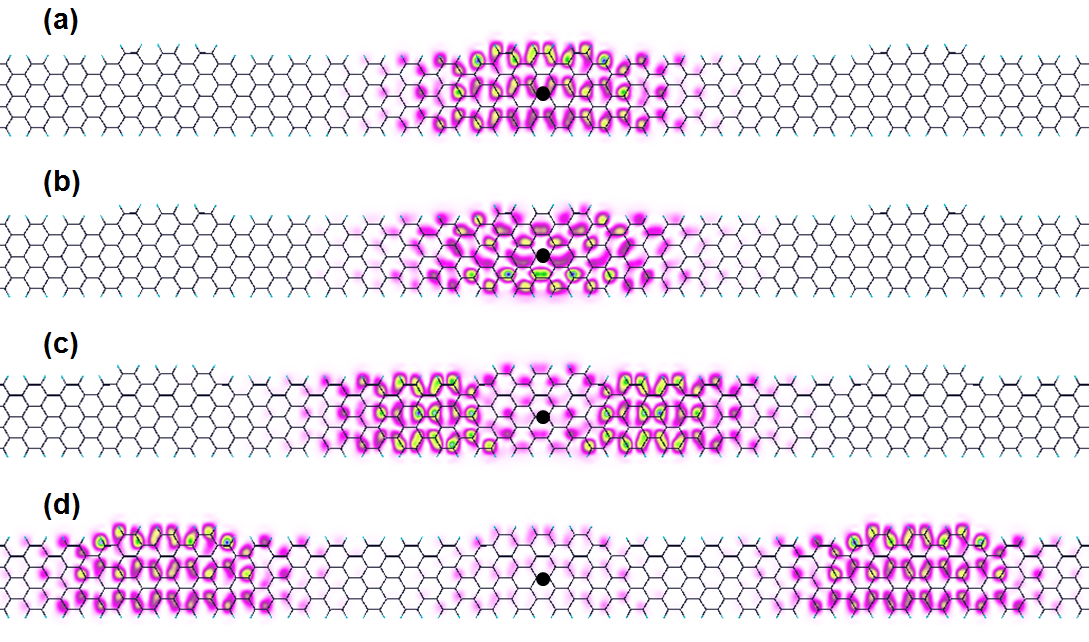}
\caption{In-plane electron distribution of the first bright (a) and few higher 
dark (b-d) excitons for the QD superlattice, where the black dot indicates 
the fixed hole.
}\label{EXCwfc2d}
\end{figure}

Let us now better focus on the nature of the first peak of the edge-modulated 
GNR. When excitonic effects are included, the first optically active excitation
still has a predominantly localized nature (i.~e. $78\%$ of the weight is 
given by transition between localized states confined in the $N=8$ region). 
Thus, the markedly different correction to the GW-RPA spectra as compared 
to ideal N=8 GNR has to be ascribed to the additional confining potential 
introduced by the edge-modulation, the first one being almost twice as large.
This change in the confinement properties is also apparent in the
excitonic wavefunction plotted along the ribbon axis (Fig.~\ref{EXCwfc1d}):
in presence of the $N=7$ barrier, the overall shape of the exciton envelope 
function changes from Gaussian (top panel) to step-like (bottom panel), with a 
significant reduction of the spatial extension and an exponential decay in the 
barrier region. Nonetheless, the exciton preserves a Wannier-like character 
[Fig.~\ref{EXCwfc2d}(a)], as found for ideal AGNRs \cite{prez+08prb}.

As described above, while the first excitation has mainly a dot-like character,
it also contains non negligible contributions coming from higher energy levels, 
thus acquiring a mixed character. This is true also for higher excitations, 
which arise from combinations of single-particle states with different 
localization properties.
For instance, the second peak is made up of several excitonic states, almost
degenerate in energy, each of them combining contributions from single
particles states localized in the dot, in the barrier, or from resonances
delocalized over the whole system. The mixed character of the excitations, as 
well as the presence of both dot- and bulk-like excitons, have to be ascribed 
to the unique nature of the system: a {\it straddling} junction between regions 
with different energy gap is here realized using the {\it same material}, 
thus preserving $\pi$-conjugation at the junction interface
\footnote{Type-II nanojuctions with peculiar optical properties can also be 
obtained from AGNRs by means of edge functionalization with appropriate 
chemical groups \cite{cocc+11jpcc,cocc+11tobe}.}, indeed different from what 
happens in common heterojunctions made of different materials 
\footnote{See e. g. BN-C nanoribbon heterostructures by \citet{sevi+08prb}}. 

%

In order to have further insights in the optical activity of such systems, we 
have investigated the presence of dark states in the low energy window for the 
periodically repeated QD structure. Figure \ref{EXCwfc2d} illustrates the 
in-plane electron distribution of the the first (bright) exciton (a) and few 
higher optically inactive excitons (b-d) of QD, plotted at fixed hole position 
(black dot). 
The edge-modulations do not introduce optically inactive states below the 
first absorption peak, inheriting the behavior of ideal ribbons 
\cite{prez+08prb}. Several dark states of distinct nature appear instead in 
the energy region between the first and the second peak 
[Fig.~\ref{EXCwfc2d}(b-d)]. 
In addition to the more common dipole forbidden states [see e.g. 
Fig.~\ref{EXCwfc2d}(b)], we find excitons which couple single-particle states 
spatially localized in different regions of the superstructure. Optical 
inactivity thus results from the small overlap of the electron and hole 
wavefunctions [see e.g. Fig.~\ref{EXCwfc2d}(c), where the hole is in the dot, 
while the electron is localized in the barrier].
A third type of exciton is shown in Fig.~\ref{EXCwfc2d}(d), where the hole 
is in the dot region and the electron is localized in its nearest replicas.

To summarize, we have investigated the electronic and optical properties of
edge-modulated AGNRs including quasi-particle corrections and excitonic 
effects. Exploiting the oscillating behavior of their energy gap, we have 
demonstrated that the simple addition (or removal) of single dimeric lines 
along the ribbon width is sufficient to obtain contextual confinement of both 
electrons and holes. We show that these nanostructures can act as optically 
active QDs, whose properties are significantly modified by many-body 
effects. Coexistence of dot-like and extended excitations, as well as fine 
tunability of optical spectra are unique features which make these systems 
conceptually different from conventional QDs. These results offer a 
tantalizing perspective, above all in light of recent production of atomically 
controlled armchair-edged GNRs.


\begin{acknowledgments}
The authors thank F. Troiani and G. Goldoni for fruitful discussions.
This work was partially funded by ``Fondazione Cassa di Risparmio di Modena'', 
Italian Ministry of Foreign Affairs, and MIUR-FIRB grant ItalNanoNet. 
CPU time was granted by CINECA through INFM-CNR.
\end{acknowledgments}



%

\end{document}